# Near-surface Extreme Wind Events and Their Responses to Climate Forcings in a Hierarchy of Global Climate Models

Gan Zhang[1], Megha Rao[1], Isla Simpson[2], Kevin A. Reed[3], Brian Medeiros[2],

Hsing-Hung Chou[4], and Tiffany Shaw[4]

[1] Department of Climate, Meteorology & Atmospheric Sciences, University of Illinois Urbana-Champaign, Urbana, Illinois

[2] NSF National Center for Atmospheric Research, Boulder, Colorado

[3] School of Marine and Atmospheric Sciences, Stony Brook University, Stony Brook, New York

[4] Department of the Geophysical Sciences, the University of Chicago, Chicago, Illinois

*Corresponding Author: Gan Zhang (gzhang13@illinois.edu)

Abstract

Near-surface extreme winds profoundly affect human society, yet process-based understanding of their changes under climate forcings remains limited. This study systematically investigates the responses of high (HWE) and low (LWE) wind extremes (10-meter) to climate forcings using a hierarchy of climate model experiments from multiple general circulation models that participated in the Cloud Feedback Model Intercomparison Project. We analyze idealized atmosphere-only aquaplanet (Aqua) simulations and more realistic land-atmosphere (AMIP) simulations to identify robust responses to climate forcings and trace the sources of structural uncertainty. In Aqua simulations, tropical LWE changes exhibit large inter-model spread, which can be traced to dynamically distinct representations of low-pressure systems between models. In contrast, extratropical HWE intensify robustly with surface warming, linked to the strengthening of high-latitude extratropical cyclones. The AMIP simulations confirm the robust intensification of extratropical HWE. The more realistic boundary conditions in AMIP simulations act as a constraint, reducing inter-model spread in tropical zonal means compared to Aqua simulations. A comparison of uniform and patterned 4-K warming experiments suggests that the global magnitude of warming, rather than the specific warming pattern, dominates the large-scale responses of wind extremes. However, regional projections of extreme wind changes, especially over land, remain highly uncertain due to divergences in model physics. Case studies reveal that major disagreements in HWE changes can stem from fundamental differences in representing the type and seasonality of extreme-producing weather systems. Our results underscore that reducing uncertainty in regional wind projections requires constraining the physical representation of weather systems in climate models.

# 1. Introduction

Near-surface wind is often observed at 10 meters above the surface and affects numerous aspects of human society, including renewable energy generation, air pollutant dispersion, and public safety. As part of the broader atmospheric circulation, near-surface winds are subject to changes driven by natural variability and anthropogenic forcings. While general circulation studies often prioritize examination of the responses of mean, free-troposphere wind to climate forcings (e.g., Simpson et al. 2014; Shaw 2019), impact-driven studies typically focus on wind extremes defined with absolute thresholds or quantile values. For example, high wind extreme (HWE) events have significant damage potential, and their statistics and physical drivers have been studied extensively (Gastineau and Soden 2009; Gentile et al. 2023; Chou et al. 2025). Conversely, low wind extreme (LWE) events (e.g., wind speed <3 m s$^{-1}$), despite their capacity to exacerbate air pollution (e.g., Elminir 2005) and diminish wind farm output (e.g., Zeng et al. 2019; Zhang 2025), have received less attention in weather and climate dynamics research. Recent studies (e.g., Shaw and Miyawaki 2024) highlighted that upper-level extreme winds accelerate at a faster rate than mean winds under climate warming, underlining the importance of a dedicated investigation into wind extremes.

The scientific community has made significant progress in exploring both the robust and uncertain aspects of changing atmospheric circulation (Shaw et al. 2024). For example, Held and Soden (2006) posited that energetic constraints would lead to a deceleration of tropical overturning circulation in warmer climates. Model simulations suggest this weakening is insensitive to the Pacific warming pattern and is already detectable in observations (Shrestha and Soden 2023). Meanwhile, successive generations of Coupled Model Intercomparison Project (CMIP) models have projected poleward shifts in midlatitude westerly jets in response to increasing $CO_2$ concentrations (e.g., Yin 2005; Lorenz and DeWeaver 2007; Simpson et al. 2014; Shaw 2019). While debates persist regarding the precise physical mechanisms (Shaw 2019; Lionello et al. 2024), significant wind changes over recent decades have become detectable in global atmospheric reanalysis (Woollings et al. 2023) and satellite observations (Di Girolamo et al. 2025). Some of the detected changes have been attributed to human activities, including greenhouse gas and aerosol forcings (Kang et al. 2024; Dong et al. 2022; Chou et al. 2025). Despite the recent progress, changes in regional circulation extremes require further investigation (Shaw et al. 2024).

Climate simulations, particularly initial-value large-ensemble simulations, have been instrumental in interpreting the observed changes amidst model biases and natural variability (Deser et al. 2020; Simpson et al. 2025). Nonetheless, even after accounting for natural variability, models exhibit profound disagreements. For instance, while climate simulations in high greenhouse gas emission scenarios predict a slowdown in the Northern Hemisphere summertime zonal-mean midlatitude circulation (e.g., Coumou et al. 2015; Dong et al. 2022; Kang et al. 2024), regional projections from multiple initial-value large-ensemble simulations suggest that changes in other seasons can be highly uncertain, especially over land regions (e.g., Zhang 2025). Such findings indicate that structural uncertainty, i.e. differences in model formulation, contribute significantly to divergent regional wind projections. Currently, few studies have investigated how this large-scale circulation uncertainty translates to potential changes in high- and low-wind extremes (Chou et al. 2025), which are often of greatest societal concern.

Bridging the gap between large-scale circulation and surface impact would benefit from a synoptic perspective. Warming-induced changes in the large-scale circulation (e.g., poleward jet shifts) and thermodynamic environment (e.g., moisture content increase) affect midlatitude storm tracks and intensify cyclones, increasing HWE in the extratropics (Gastineau and Soden 2009;

Priestley and Catto 2022; Gentile et al. 2023). Beyond the midlatitudes, circulation and environmental changes modulate the frequency and spatial-temporal distribution of extreme-generating weather systems, such as tropical cyclones (e.g., Sharmila and Walsh 2018; Zhang 2023). However, existing studies focus on wind extremes in disconnected regions, leaving fundamental questions concerning wind extremes and their associations with weather systems unanswered. For example, the fidelity and diversity among climate models in representing the weather systems driving HWE and LWE across regions, and their response to climate forcing, remain largely unassessed. Narrowing the uncertainty in regional projections requires addressing these foundational questions.

A systematic application of a model hierarchy approach (Held 2005; Maher et al. 2019) to a multi-model ensemble offers a valuable strategy for highlighting robust responses and identifying primary uncertainty sources. For example, Stevens and Bony (2013) analyzed multi-model Aqua simulations forced with identical climate forcings and highlighted vastly different responses of the Intertropical Convergence Zone (ITCZ), directly attributable to differences in the formulations of atmospheric models. Medeiros et al. (2015) expanded on that to show robust responses in the large-scale circulation to uniform warming but different climate feedbacks, further bolstering the idea that the structural uncertainty is critical for the climate response. By pairing idealized aquaplanet (Aqua) simulations with more realistic atmospheric model intercomparison project (AMIP) simulations, a hierarchy of models can help investigate physical mechanisms and disentangle uncertainty sources affecting global wind extremes. Here we work with hierarchy experiments conducted under the auspices of the Cloud Feedback Model Intercomparison Project (CFMIP) phase 3 (Webb et al. 2017), which is a specialized, endorsed project within the Coupled Model Intercomparison Project (CMIP) Phase 6.

The CFMIP also provides a unique opportunity to dissect the impacts of sea surface temperature (SST) warming on extreme wind events. Past studies suggest that the impacts of uniform and patterned warming are comparable for precipitation responses (Ma and Xie 2013). Recent studies also emphasized the substantial impacts of SST warming pattern on the Earth's equilibrium climate sensitivity (e.g., Sherwood et al. 2020 and references therein) and high-impact weather systems (Sobel et al. 2023; Zhao and Knutson 2024). In comparison, the responses of some large-scale circulation features and atmospheric blocking are dominated by uniform warming rather than the pattern of warming (Ma and Xie 2013; Shrestha and Soden 2023; Narinesingh et al. 2024). Furthermore, the uniform and the patterned warming can each dominate the responses of regional circulation in different months (e.g., Chadwick et al. 2026). A comparison of uniform and patterned warming simulations from CFMIP is thus valuable for determining whether wind extremes are influenced by patterned warming versus the magnitude of the warming.

Building upon the theoretical and modeling foundation, we leverage the CFMIP model hierarchy to examine HWE and LWE events in the atmosphere-only Aqua simulations and land-atmosphere AMIP simulations. We analyze the responses of selected models (Section 2) to identical climate forcings, including quadrupling $CO_2$ concentrations, uniform 4K warming, and patterned 4K warming. We acknowledge the inherent limitations related to the lack of coupled atmosphere-ocean dynamics and relatively low resolutions of CFMIP simulations. Our aim here is to isolate atmospheric responses and understand them in this simplified setting. We hypothesize that the responses of mean wind and extreme wind differ from each other. Our analysis seeks to answer the following research questions:

- How do $CO_2$ forcing and SST warming affect the extremes of near-surface wind speed in the simulations across the model hierarchy?

- How do Aqua and AMIP simulations differ in their representations of extreme wind responses to diverse climate forcings?
- Can the simulated changes in near-surface wind extremes and their disagreements be directly linked to synoptic-scale weather systems?

The rest of the study is organized as follows: Section 2 details the Data and Methodology employed. Section 3 presents the analysis of changes in mean and extreme wind within Aqua simulations and examines their association with synoptic weather systems. Section 4 focuses on AMIP simulations and provides a comparative analysis with Aqua results. Finally, Section 5 summarizes our findings and discusses their implications for future research.

## 2. Data and Methodology

### 2.1 Model Simulation Data

This study analyzes CFMIP phase 3 simulations (Webb et al. 2017) from Aqua (10 years in length) and AMIP experiments (1979-2014). Each set of experiments were conducted with multiple models driven by identical boundary forcings. The Aqua experiments utilize atmosphere-only models forced by a zonally symmetric, analytical SST pattern (Neale and Hoskins 2000), and are devoid of a seasonal cycle, sea ice, or other surface complications. The insolation is set to a solar constant of 1365 W m$^{-2}$ with a diurnal cycle and stays in perpetual equinox conditions. The Aqua experiments include a control run ($CO_2$=348 ppmv), a quadrupling $CO_2$ (4×$CO_2$) experiment, and a uniform 4K warming (P4K) experiment. In contrast, the AMIP experiments employ atmosphere and land models with prescribed SST with observation-based patterns. These simulations include realistic land-atmosphere interactions and zonally asymmetric boundary conditions. The AMIP boundary forcings include preindustrial control (pi-control), 4×$CO_2$, P4K, and Future4K experiments. The Future4K experiments use the composite SST warming pattern derived from CMIP3, scaled to a global mean warming of 4K over ice-free oceans. Comprehensive details regarding the boundary forcings are documented by Webb et al. (2017). The Aqua and AMIP simulations with different models help isolate the impacts of model formulation differences (with identical forcings) and boundary forcings (with identical models).

We analyze simulations from six CFMIP models (Table 1). While data availability issues prevented the analysis of more models, our selection here still captures a range of dynamical behaviors. The synoptic-scale analysis primarily focuses on the outputs from the Community Earth System Model v2 (CESM2) (Danabasoglu et al. 2020) and the IPSL-CM6A-LR (Boucher et al. 2020). CESM2 serves as a primary reference due to its widespread lineage (Kuma et al. 2023), while IPSL-CM6A-LR was selected to provide a dynamical contrast, specifically due to its simulation of weaker tropical wave activity (Medeiros et al. 2021). TaiESM1 (Lee et al. 2020) is included to examine structural sensitivity within the same model family as CESM2. The analysis focuses on the last 10 years of the Aqua simulations and the last 30 years (1985-2014) of the AMIP simulations. We note that the native grid spacing of all the examined models (>1 latitude/longitude degree) limits the resolution of wind extremes associated with fine-scale features, a limitation discussed further in Section 5.

**Table 1 Model configuration and information.** The "Notes" column indicates lineage relationships (e.g., TaiESM1 and CESM2) or specific models selected for the detailed Lagrangian studies in Sections 3 and 4 (IPSL-CM6A-LR).

|  | **Model Reference** | **Grid (lat, lon)** | **Notes** |
|---|---|---|---|
| **CESM2** | (Danabasoglu et al. 2020) | 192, 288 | Primary model for synoptic analysis |
| **CNRM-CM6-1** | (Voldoire et al. 2019) | 128, 256 | |
| **HadGEM3-GC31-LL** | (Andrews et al. 2020) | 144, 192 | |
| **IPSL-CM6A-LR** | (Boucher et al. 2020) | 143, 144 | Selected for comparative study with CESM2 (Sec. 2.2) |
| **MRI-ESM2-0** | (Yukimoto et al. 2019) | 160, 320 | |
| **TaiESM1** | (Lee et al. 2020) | 192, 288 | Shares code lineage with CESM2 |

## 2.2 Analytics Method

We examine the daily 10-meter near-surface wind speed using both Eulerian and Lagrangian approaches. The Eulerian analysis characterizes large-scale changes and is performed on the native grid of the CFMIP simulation output. Within these analysis periods of Aqua and AMIP simulations (Section 2.1), we calculate the mean, the $1^{st}$ (LWE), and $99^{th}$ (HWE) percentiles of 10-meter wind speed for each grid point. Although changes in wind means and extremes in climate simulations exhibit strong seasonality (Zhang 2025; Chou et al. 2025), we follow prior studies (e.g., Gastineau and Soden 2009) and do not impose seasonally dependent percentile thresholds. This choice maintains consistency between season-less Aqua and seasonal AMIP simulations. The wind speeds corresponding to the percentile values are analyzed in latitude-longitude coordinates to highlight the intensity change in the events of the same likelihood (i.e., $1^{st}$ and $99^{th}$ percentiles). Additionally, we evaluate zonal averages to help interested readers compare the findings with zonal-mean analyses in previous studies.

To link wind extremes to specific synoptic-scale weather features, we perform a Lagrangian feature-tracking analysis on CESM2 and IPSL-CM6A-LR outputs (Section 2.1). A comparison of the two models helps develop a process-based understanding of structural uncertainties related to model configuration including physics formulation and grid resolution. For each experiment by the two models, we use the TempestExtremes package (Ullrich et al. 2021) to identify contiguous regions that meet the wind speed thresholds for LWE ($1^{st}$ percentile) and HWE ($99^{th}$ percentile). Consistent with the Eulerian analysis, the thresholds are calculated for each grid point. The grid points within these flagged regions are consolidated into individual objects and archived for subsequent analysis. Crucially, no minimum size threshold is applied to the flagged contiguous regions, ensuring that both highly localized and broad extreme events are captured in the object consolidation. More details about the feature processing are available in the TempestExtremes documentation.

We analyze objects of HWE and LWE in selected regions to highlight process-dependent uncertainties. The analyses of the Aqua simulations focus on objects with their centroids located within 90°E-270°E in the Northern Hemisphere. The choice helps reduce the data volume for analysis, and the specific longitudinal range and hemisphere choice does not affect the generalizability of our findings in zonally and hemispherically symmetric simulations. Because our Lagrangian analysis covers regions >1000 km away from the object centroids, a latitudinal limit (10°N-65°N) is implemented to mitigate issues related to the polar singularity and cross-equator weather systems. The objects are grouped based on their centroid latitude to analyze weather systems in the tropics (10°N-25°N) and on the equatorward (30°N-43°N) and poleward (50°N-65°N) flanks of the midlatitude jet. The latitudinal edges are chosen to address the uncertainty of the center latitude of the subtropical high and the midlatitude westerly simulated by CFMIP models. Small changes up to a few degrees do not qualitatively change our findings (not shown). The analyses of the AMIP simulations focus on objects in the Northwest Pacific region (10°N-30°N, 120°E-150°E), a region known for high frequency synoptic variability, strong land-sea contrast, and where initial analysis revealed significant inter-model disagreement in extreme wind responses (Section 4).

Within each of these selected regions, we use the geometric centroid of each HWE and LWE object to construct the centroid-centered composites in relative latitude-longitude coordinates. For each event object, we apply a stereographic projection to the nearby data and analyze the domain within a 2000 km radius. To avoid distance mismatch related to the original latitude-longitude grid, we calculate anomalies relative to the local climatology and regrid the original simulation data to polar coordinates before constructing composites. For each model, we examine the differences between samples from control and perturbed experiments. Their statistical significance is evaluated with two-sided Student's t-test, assuming that the large sample size ($N \sim 10^3 - 10^4$) ensures robust results despite potential serial correlation that reduces the size of independent samples.

## 3. Aqua Simulations

### 3.1 Zonal Means of Wind Means and Extremes

The Aqua control simulations from the six models produced similar zonal average structure of mean and extreme wind (Figure 1a-c). The mean wind shows midlatitude peaks associated with the storm tracks near 45 degrees and tropical peaks associated with the trade winds (Figure 1a). The low values in the subtropics and near the equator correspond to subtropical highs and the ITCZ, respectively. The wind speed of LWE events follows a similar pattern (Figure 1b), while its HWE counterpart (Figure 1c) shows more prominent high values associated with midlatitude storm tracks. The model spread among the three metrics are up to ~2 m s$^{-1}$, with CESM2 and TaiESM1 producing higher wind speed values. Other models, such as MRI-ESM2-0 and CNRM-CM6-1, also show relatively high wind speed at some latitudes. While CESM2 and TaiESM1 possess relatively high spatial resolution (Table 1), the lack of a strict correlation with resolution across all models and metrics suggests that differences in physical parameterizations (e.g., boundary layer physics) and physics-dynamics coupling contribute significantly to the spread.

While all models simulate a robust poleward shift of the midlatitude westerlies in response to SST warming, their projections of 10-meter wind speed reveal significant inter-model spread, with inter-model spread often comparable to the magnitude of the forced response itself (Figures 1a, d, g). The mean wind response to increased $CO_2$ with fixed SST is typically small (< 0.5 m s$^{-1}$) (Figure 1d), whereas the response to uniform warming can exceed 1 m s$^{-1}$ (Figure 1g). Nearly all the models simulate a poleward shift of the midlatitude jet near 45 degrees. However, in both the midlatitudes and tropics, differences between models can reach 0.5 to 1 m s$^{-1}$, suggesting that variations from model formulation are comparable to the responses to the identical, strong climate forcings. Some of the spread may arise from the internal variability in the relatively short 10-year simulations, though the relatively homogenous responses on the latitude-longitude grid (not shown) suggest the impact of internal variability is likely secondary. In the tropics, the large model-related uncertainties for near-surface wind may be related to the profound inter-model disagreements in tropical precipitation and ITCZ width highlighted by Stevens and Bony (2013).

The zonal mean LWE and its changes (Figures 1b, e, h) largely mirror the mean wind but are characterized by even higher inter-model spreads. In regions dominated by midlatitude westerlies and trade winds, the LWE in the control experiments generally shows relatively high values of wind speed (>3 m s$^{-1}$) (Figure 1b). Responses to increased $CO_2$ forcing alone are weak (generally <0.25 m s$^{-1}$), though some models simulate larger values in association with poleward jet shifts. Responses to uniform 4K warming are even larger and occur in conjunction with the poleward shift of the midlatitude westerlies. Interestingly, a large spread (~2 m s$^{-1}$) of simulated LWE responses is evident in the trade wind region, similar to the finding by Stevens and Bony (2013). Here, CESM2 and TaiESM1 suggest weakening of LWE, while IPSL-CM6A and other models suggest the opposite. The profound disagreement in the tropics (10°N-25°N) may be linked to differing model representations of tropical low-pressure systems (Section 3.3).

The zonal-mean HWE (Figure 1c, f, i) shows a pronounced increase in high-latitude wind speed in the 4K warming experiments. In the control climatology, the zonal mean HWE show midlatitude peaks collocated with the mean westerlies near 45 ºN and 45 ºS. Secondary peaks in the tropics are about 4 m s$^{-1}$ stronger than the mean value (Figure 1a). Responses to increased $CO_2$ reach up to 0.5 m s$^{-1}$ in the high-latitude region but are notably weaker than responses to the 4K warming experiments. Under 4K warming, nearly all the models simulate an increase of 0.5 to 1 m s$^{-1}$ in HWE poleward of 45 degrees. Crucially, this response is substantially larger than the change in the mean wind, suggesting that the changes are not caused by a simple poleward shift of the entire wind distribution but rather by a strengthening of the most intense wind events on the poleward flank of the jet.

An analysis of statistical distributions of wind speed across latitudes highlights varied responses across quantiles (Figure 2). Focusing on CESM2 and IPSL-CM6A-LR, two models near opposite ends of the spread envelope (Figure 1), the distributions of wind speed change at different paces in response to climate forcings. This difference is particularly notable for the responses to the uniform 4K warming (P4K) (Figure 2c and 2f). Between 10ºS-10ºN and away from the equator, CESM2 suggests the low-quantile wind decreases at a faster pace, whereas IPSL-CM6A-LR suggests the high-quantile wind experience a more pronounced weakening. In contrast to the disagreement, both models suggest high-quantile wind accelerates faster on the poleward flank of the midlatitude jet. This high-latitude consistency appears consistent with the "fast-gets-faster" paradigm of the upper-tropospheric wind in response to warming (Shaw and Miyawaki 2024).

Collectively, these results suggest the poleward shift in the mean westerlies is closely associated with concurrent intensity changes in LWE and HWE. Crucially, the near-surface HWE (10-meter) strengthens faster, especially at high-latitudes. Furthermore, responses to uniform warming are stronger than those induced with the direct effects of $CO_2$ forcing with fixed surface temperature. This contrast appears consistent with theoretical expectations regarding the changes in static stability and latent heat release forced by surface warming (e.g., Pfahl et al. 2015; Shaw 2019).

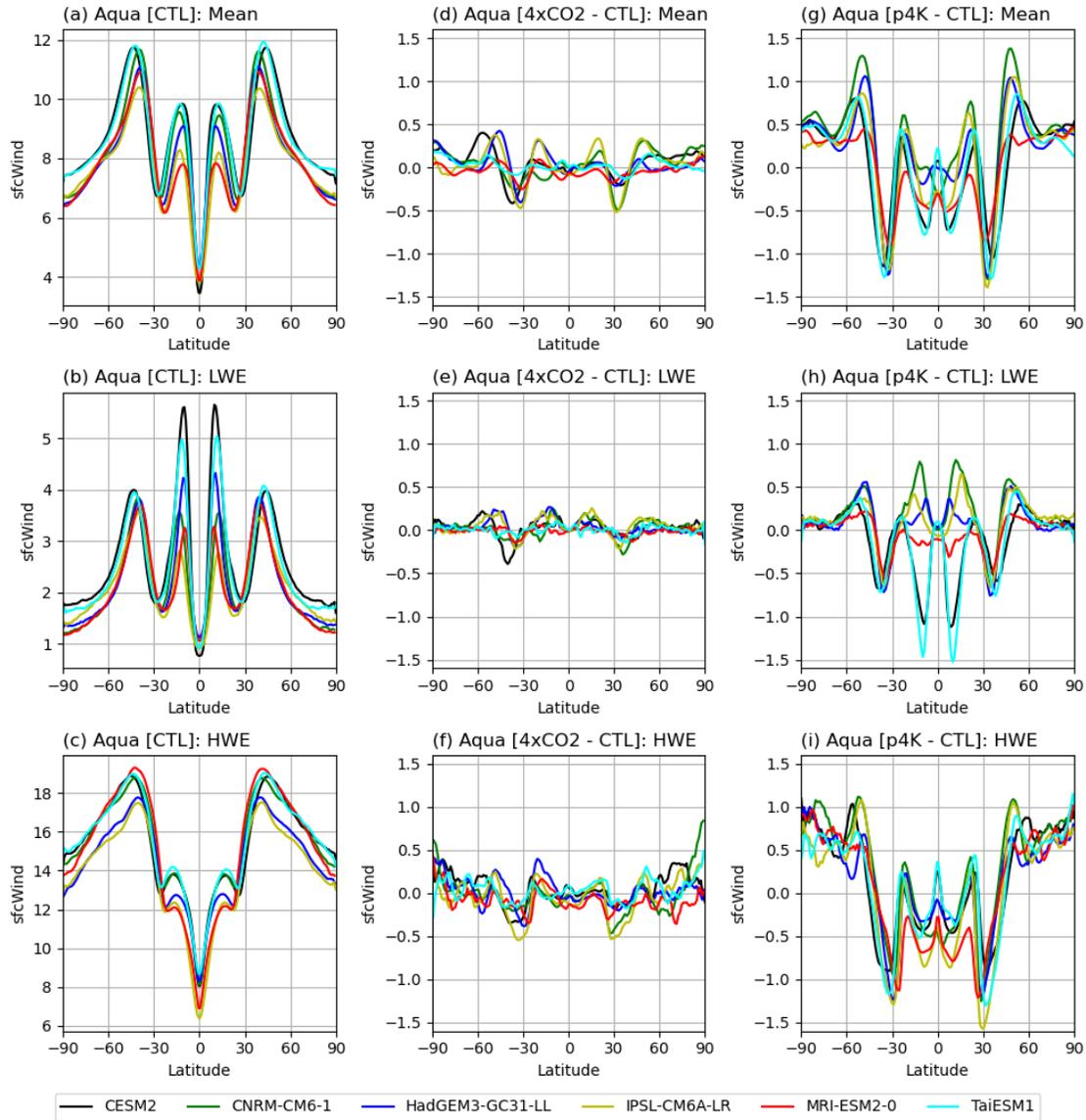

Figure 1. Zonal mean wind speed (m s$^{-1}$) of 10-year means, LWE, and HWE events in Aqua experiments. (a) Mean wind speed in Aqua control experiments. (b) LWE (1$^{st}$-percentile) wind speed in Aqua control experiments. (c) HWE (99$^{th}$-percentile) wind speed in Aqua control experiments. (d-f) Same as (a-c), but for the differences between Aqua 4×CO$_2$ and control experiments. (g-i) Same as (a-c), but for the differences between P4K and control experiments. The values of LWE and HWE are first evaluated on each grid point using daily 10-meter wind speed data, and zonal averaging is applied after the quantile calculation.

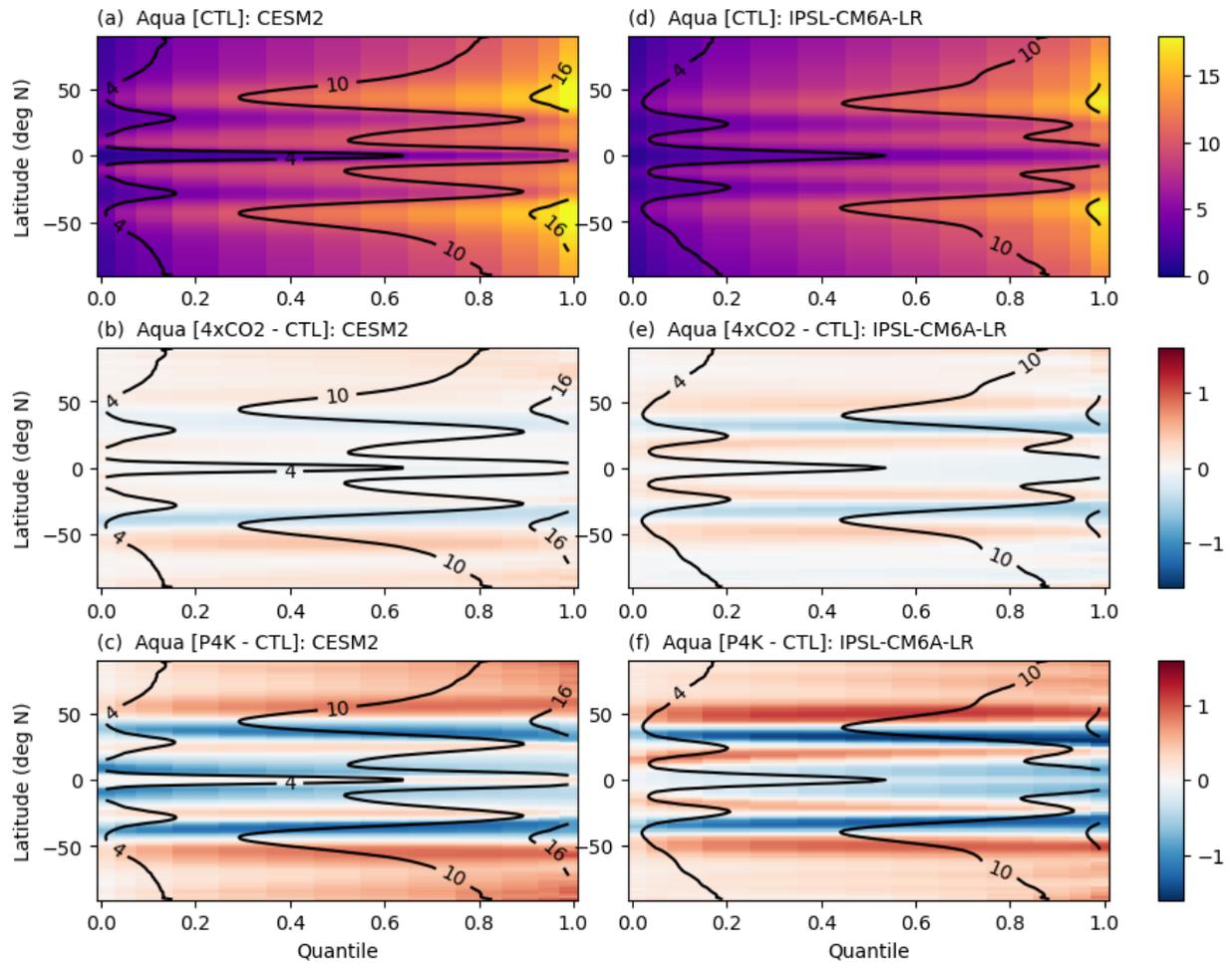

Figure 2 Zonal averages of the quantile values (0.01 to 0.99) of 10-meter wind speed at each latitude. (a) Quantile-latitude values in CESM2 Aqua control experiment. The shading shows the rasterized field, and the contours show the smoothed field. Between the quantile tails (0.01 and 0.99), the bin interval is 0.05 between 0.05 and 0.95. (b) Differences in the quantile-latitude values between CESM2 Aqua 4×$CO_2$ and control experiments (shading). The contours show the reference values from Aqua control experiment. (c) Same as (b), but for the differences between P4K and control experiments. (d-f) Same as (a-c), but for IPSL-CM6A-LR experiments.

## 3.2 Changes in Extreme-generating Weather Systems

To understand the physical mechanisms behind the zonal mean changes (Section 3.1), we composite the synoptic-scale features associated with LWE and HWE events. Based on the latitude-dependent wind responses in Figure 1, we analyze events within three distinct latitude bands: jet-poleward (50°N-65°N), jet-equatorward (30°N-43°N), and tropics (10°N-25°N). The separation of synoptic systems on the jet flanks accounts for the impacts of meridional wind shear on their growth rate and maximum intensity (Hadas and Kaspi 2025). We first examine CESM2 to establish a baseline representation of synoptic features in a model with standard tropical wave activity, before contrasting it with IPSL-CM6A-LR to diagnose sources of model spreads (Medeiros et al. 2021).

LWE in the CESM2 simulations are linked to extratropical high-pressure anomalies (Figures 3a and 3d) and tropical low-pressure anomalies (Figure 3g). These systems are characterized by relatively weak meridional pressure gradients. Consistent with the findings from Eulerian means (Figure 1), the responses of these systems to increased $CO_2$ forcing are generally weak with the fixed SST. Under the uniform P4K warming, the strengthening of an anomalous pressure dipole reduces meridional pressure gradients near LWE on the poleward flank of the midlatitude jet (Figure 3c), partly offsetting the increase of mean wind speed (Figure 1g) for the LWE events (Figure 1h). Near LWE on the equatorward flank of the midlatitude jet, high-pressure anomalies weaken the meridional pressure gradient (Figure 3d). The anomalies weaken in response to 4K warming and partly offset the increase of mean wind (Figure 1g). In the tropics, 4K warming strengthens low-pressure anomalies (Figure 3g and 3i), weakening the pressure gradient in the subtropical-high region. The associated reduction of wind speed is broad (Figure 3i) and suggest an increase in the spatial extent of LWE events in a warmer climate.

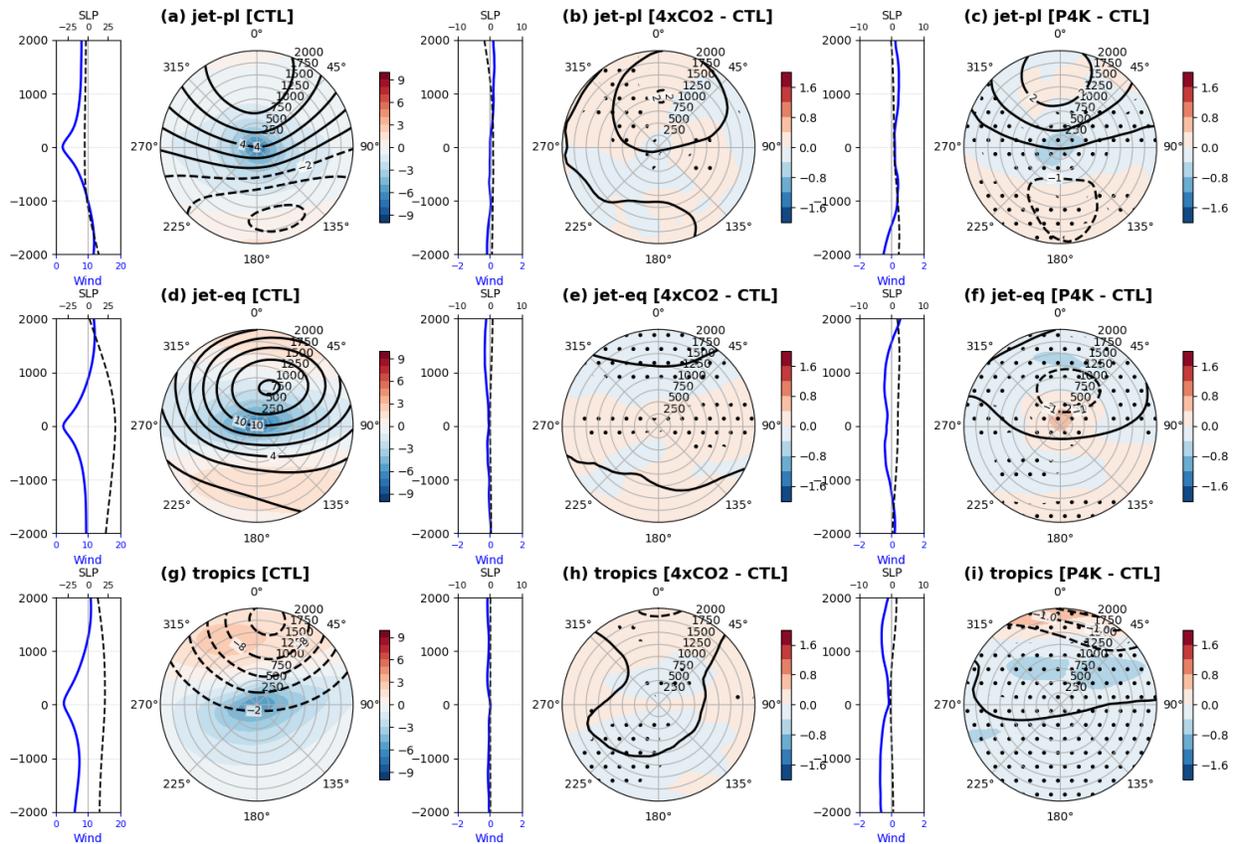

*Figure 3. Composites of LWE events in CESM2 experiments. (a) Right side show anomalies of 10-meter wind speed (shading; m s$^{-1}$) and sea level pressure (contours; hPa). The polar coordinate is centered at the centroids of identified LWE events identified on the poleward flank of the midlatitude jet (50°N-65°N). The top corresponds to the North, and the radius is denoted in km. The left shows the profile of wind speed (blue solid) and sea level pressure (black dashed). For clarity, sea level pressure field shows the difference of original values and 1000 hPa. (b) Same as (a), but for the differences between Aqua 4×CO$_2$ and control experiments. The stippling highlights significant differences at the 95% confidence level. (c) Same as (a), but for the differences between P4K and control experiments. (d-f) Same as (a-c), but for the LWE events on the equatorward flank of the midlatitude jet (30°N-43°N). (g-i) Same as (a-c), but for the LWE events in the tropics (10°N-25°N).*

Conversely, HWE events are consistently associated with synoptic systems that generate strong pressure gradients, such as intense surface cyclones (Figure 4a and 4d) and subtropical anticyclones (Figure 4g). In the extratropics, the HWE events often appear near the centers of surface cyclones. These systems respond weakly to the $CO_2$ forcings except on the poleward flank of the midlatitude jet. Under the P4K warming, the pressure anomalies of extratropical cyclones decrease and strengthen HWE. The intensification of extratropical cyclones is consistent with theoretical and modeling studies, which link cyclone intensification to enhanced latent heat release in a warmer, moister atmosphere (e.g., Pfahl et al. 2015; Tierney et al. 2018; Tamarin and Kaspi 2017). Additionally, the wind intensification and pressure deepening under the P4K warming occurs over broad regions near the geometric centers of HWE and extratropical cyclones. This suggests an expansion of high-wind regions and associated cyclones, which is qualitatively consistent with the expansion of the free-troposphere circulation of extratropical cyclones (Dai and Nie 2022).

The intensification of extratropical cyclones and HWE is consistent with the changes in the Eulerian means on the poleward flank of the midlatitude jet (50°N-65°N) but not on the equatorward flank (30°N-43°N) (Figure 1i). However, re-evaluating the HWE events using full wind fields confirms that the storm-centered intensification is consistent with the Eulerian changes, including the HWE weakening on the equatorward flank (not shown). In this region, the warming-induced poleward shift of the midlatitude storm track reduces the frequency of traversing extratropical cyclones (Schemm and Röthlisberger 2024). This frequency-driven reduction dominates the Eulerian signal of HWE changes, masking the underlying behavior of individual systems whose isolated intensities slightly increased (Figure 4f). Further south, pressure changes weaken the anticyclone near the centroid of HWE (Figure 4j), contributing to the weakening of HWE consistent with the Eulerian mean changes. Overall, the latitude-dependent response highlights the complexity of projecting future changes in wind extremes, as research of a single class of features (e.g., extratropical cyclones) or a single feature property (e.g., intensity) cannot fully explain the details of projected wind changes.

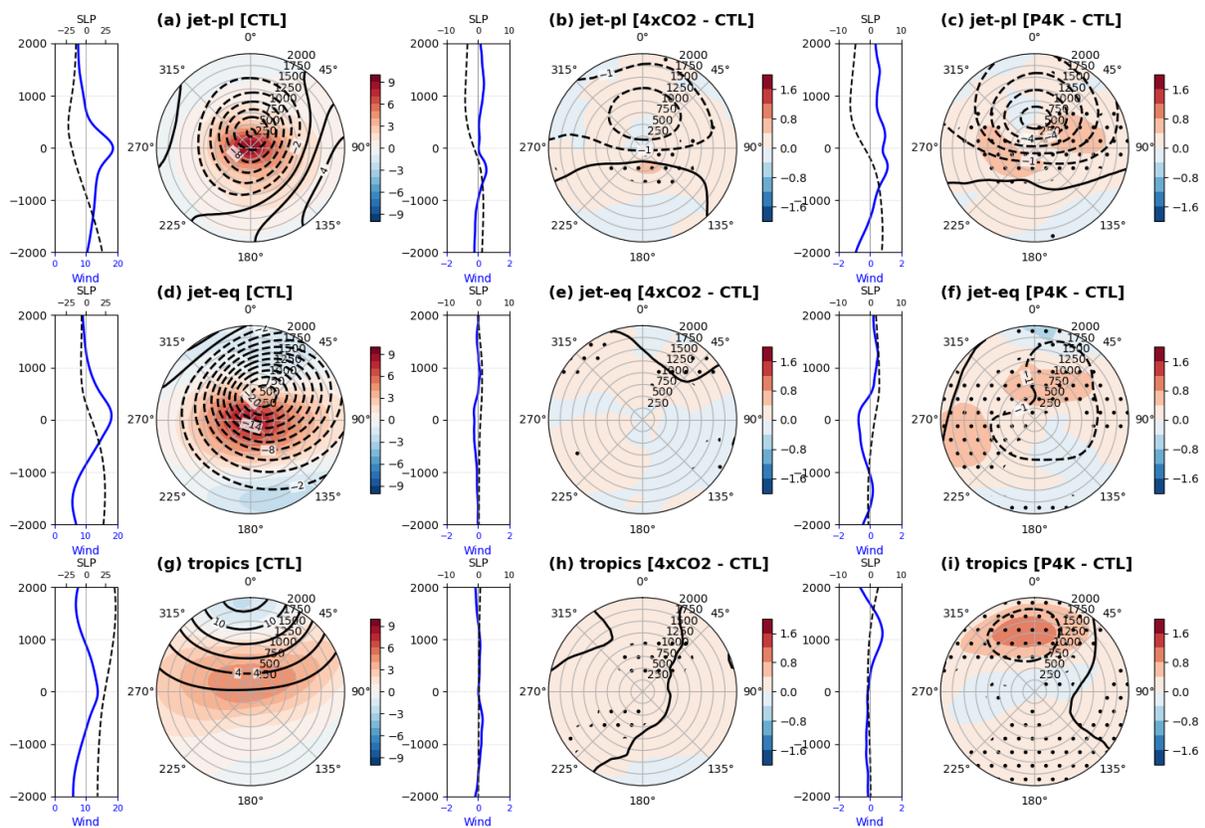

*Figure 4. Composites of HWE events in CESM2 experiments. The other plot settings are the same as Figure 3.*

This composite analysis provides insights for diagnosing the source of inter-model spread. Despite large differences in the Eulerian changes projected by CESM2 and IPSL-CM6A-LR, the two models generally suggest wind extremes are associated with similar types of weather systems across latitudes. However, the simulated responses to identical climate forcings differ in magnitude. Interestingly, even though the IPSL-CM6A-LR has coarser grid spacing (Table 1), the model simulates much stronger responses of surface cyclones and anticyclones to climate forcings. The differences in wind responses suggest that the magnitude of wind extreme changes simulated by a climate model do not necessarily depend on the model resolution alone. Finally, the simulated changes (e.g., Figure 5c and Figure 6c) suggest that some extreme-generating weather systems on the poleward flank of the midlatitude jet may experience notable shifts or expansions.

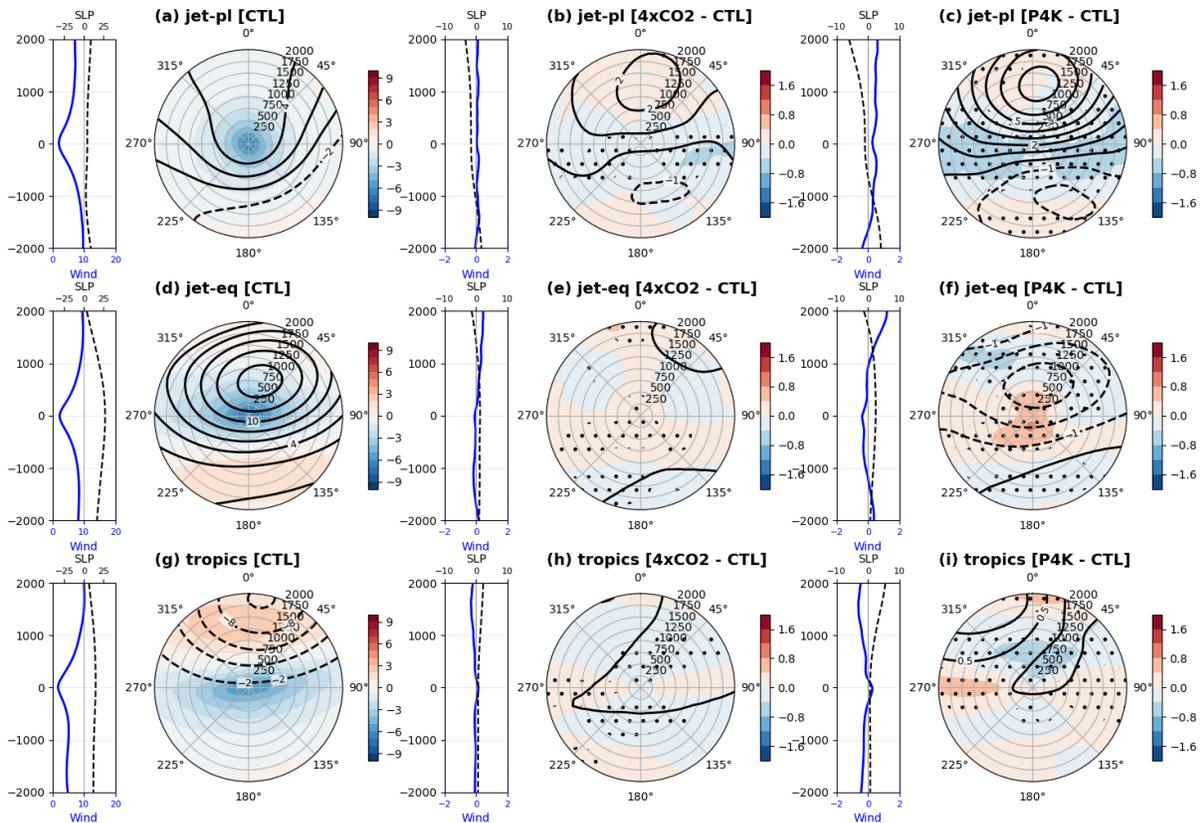

*Figure 5. Composites of LWE events in IPSL-CM6A-LR experiments. The other plot settings are the same as Figure 3.*

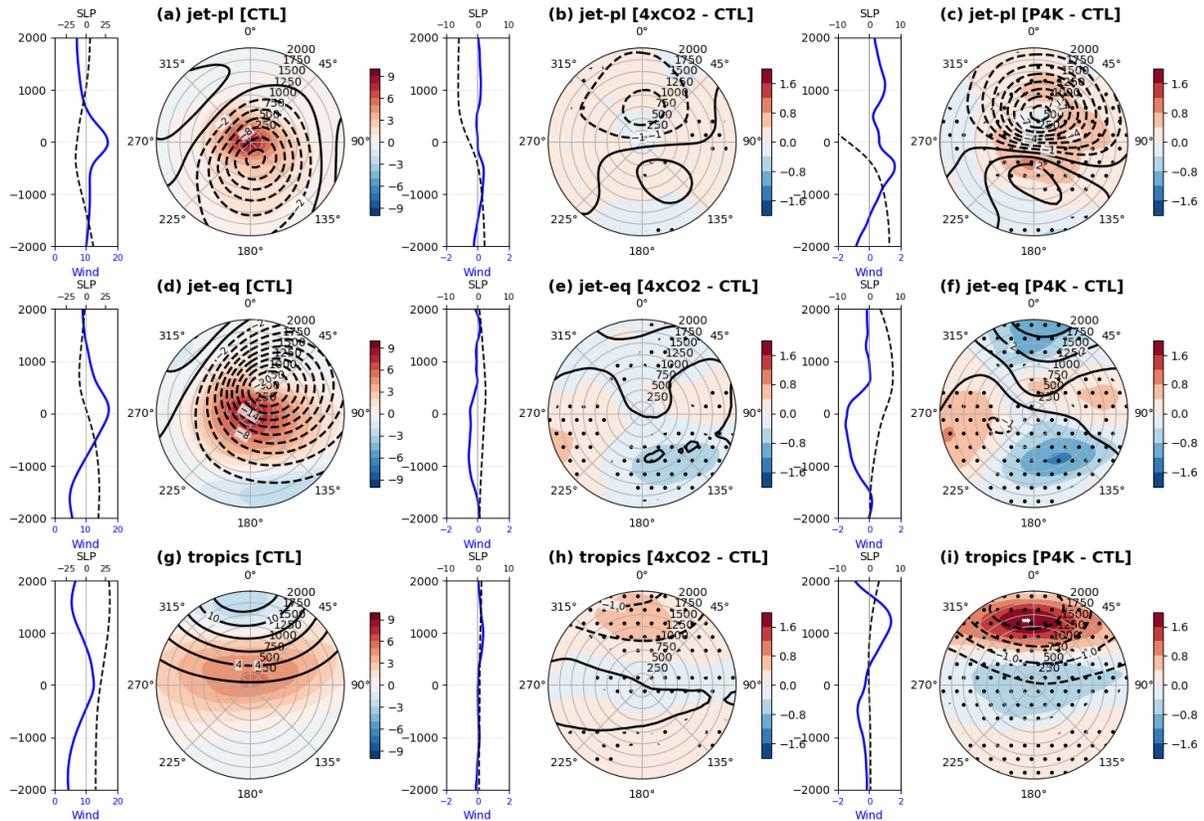

*Figure 6. Composites of HWE events in IPSL-CM6A-LR experiments. The other plot settings are the same as Figure 3.*

## 4. AMIP Simulations

We now transition to the AMIP simulations to assess how the insights from the idealized aquaplanet setup translate to a more realistic configuration with zonally asymmetric boundary forcings. Our analysis focuses on the consistency between Aqua and AMIP responses, the sensitivity to the SST warming pattern, and the sources of tropical spread in wind responses.

## 4.1 Zonal Means of Wind Means and Extremes

Comparing the AMIP simulations (Figure 7) to their Aqua counterparts, the zonal-mean responses of mean and extreme wind speeds show notable hemispheric asymmetry. The most striking changes occur in the southern hemisphere and resemble the poleward shift of midlatitude westerlies simulated by the Aqua simulations. The magnitude of the Southern Hemisphere responses to increased $CO_2$ and uniform 4K warming is also comparable between Aqua and AMIP simulations. This similarity of responses is commensurate with the relative zonal symmetry of the southern hemisphere. Conversely, the zonal mean changes in the Northern Hemisphere are nearly zero in AMIP simulations, except for wind speed increases at high latitudes. The consistency in high-latitude wind responses across Aqua and AMIP experiments suggests these responses are relatively insensitive to surface boundary properties, including but not limited to SST patterns.

Compared to the Aqua simulation, the zonal-mean changes of LWE in AMIP simulations exhibit greater consistency among models, particularly in the tropics (Figures 7e, h, k). While the larger sample size in AMIP simulations (30 years vs. 10 years) may have contributed to statistical robustness, the greater agreement on the sign and magnitude of the tropical LWE response (Figure 1h) suggests that realistic boundary conditions act as a strong constraint, reducing inter-model spread. In other words, the responses of LWE in the tropics are likely highly sensitive to the surface boundary conditions, at least in the examined models. Whether this strong constraint by land represents the consequences of accurately capturing real-world physics or the influence of tuning parameterized physics (including potentially compensating errors) around the real-world state is unclear.

In the AMIP simulations, the response of HWE is also characterized by intensification of HWE at high latitudes (Figures 7I, l). In the Southern Hemisphere, the high-latitude intensification and the HWE displacement in the midlatitudes are broadly similar to the Aqua simulations. While the Northern Hemisphere responses are notably weaker, the models also suggest general intensification of HWE. The intensification of high-latitude HWE can be linked to the intensification of extratropical cyclones (Section 4.2). A direct comparison of the intensification magnitude with previous studies (Priestley and Catto 2022; Gentile et al. 2023) is challenging due to model and method differences. For example, the magnitude of high-latitude changes simulated by CFMIP phase 3 models appears comparable to that of CMIP6 models with Shared Socioeconomic Pathway (SSP) 5-8.5 forcing (Priestley and Catto 2022). However, the magnitude appears substantially weaker than the changes simulated by a high-resolution (50-km) climate model forced with 2-K uniform warming (Gentile et al. 2023).

Another key finding is that the zonal-mean responses of mean and extreme winds is largely insensitive to the precise spatial structure of the warming. The changes in the uniform warming (P4K; Figures 7g, h, i) and patterned warming (Future4K; Figures 7j, k, l) experiments are remarkably similar, suggesting that for the zonal-mean changes of wind means and extremes, the magnitude of SST warming is a more dominant driver of large-scale signals than its regional pattern. Furthermore, the Southern Hemisphere responses across both the idealized Aqua and realistic AMIP frameworks are highly consistent. This pattern-insensitivity differs from the sensitivity of tropical precipitation in the 21$^{st}$ century simulations and 4-K warming experiments (Ma and Xie 2013) but is qualitatively consistent with studies on the large-scale circulation and dynamical features (Medeiros et al. 2015; Shrestha and Soden 2023; Narinesingh et al. 2024). An analysis of quantile-latitude distributions (not shown) suggests the warming pattern can introduce some differences in the tropics, though the pattern-associated spread appears less important than the spread associated with model structural uncertainty. Our findings about global wind extremes complement the relatively few studies on the sensitivity of extremes to SST warming patterns (e.g., Sobel et al. 2023).

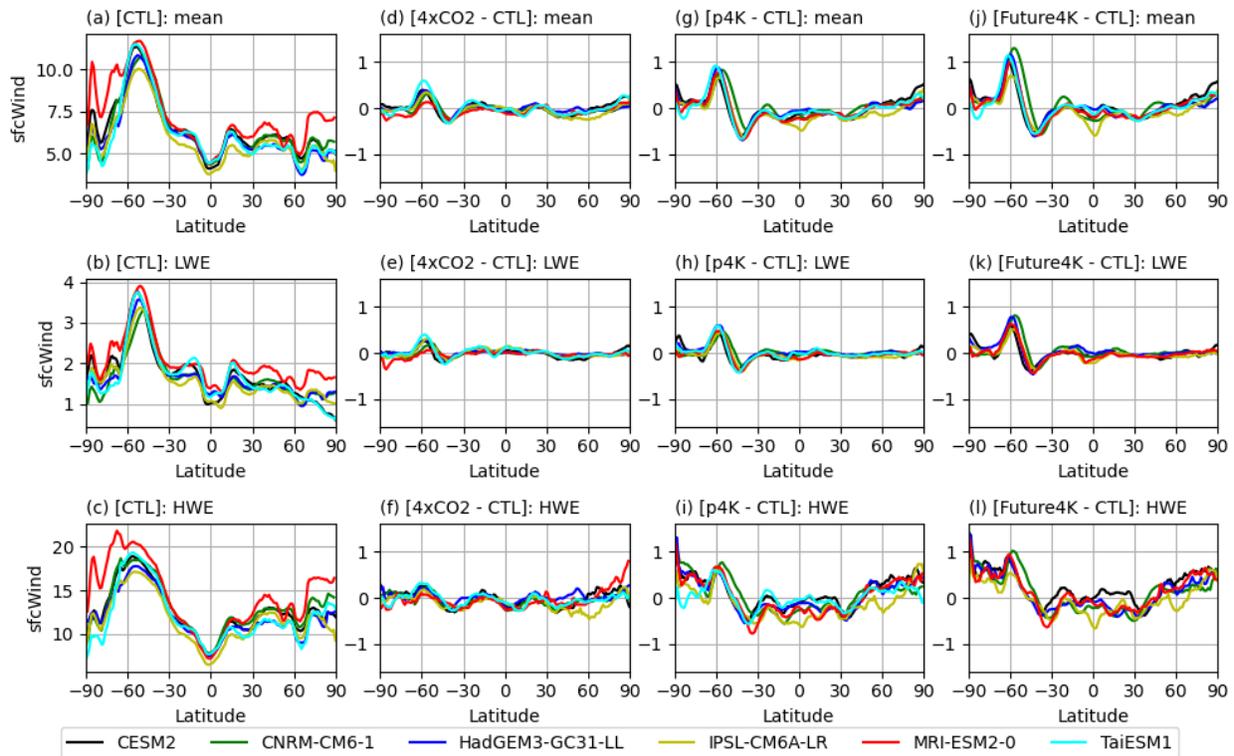

Figure 7. Zonal mean wind speed (m s$^{-1}$) of 30-year means, LWE, and HWE events in AMIP experiments. (a) Mean wind speed in Aqua pi-control experiments. (b) LWE (1$^{st}$-percentile) wind speed in AMIP pi-control experiments. (c) HWE (99$^{th}$-percentile) wind speed in AMIP pi-control experiments. (d-f) Same as (a-c), but for the differences between $4 \times CO_2$ and pi-control experiments. (g-i) Same as (a-c), but for the differences between P4K and pi-control experiments. (j-l) Same as (a-c), but for the differences between Future4K and pi-control experiments. Panels k and l do not include TaiESM1 due to the lack of daily surface wind data.

## 4.2 Regional Uncertainties and Their Sources

Since the wind responses in 4×$CO_2$ are generally small (Figure 7), here we focus on the regional wind responses in P4K and Future4K experiments (Figures 8-10). Compared to the zonal-mean responses, regional responses in the AMIP simulations are more complex and model-dependent. The large-scale circulation changes can be interpreted as responses to SST forcings modulated by land-sea contrast and interactions with stationary waves (e.g., Shaw and Voigt 2015; Wills et al. 2019).

Over the extratropical oceans, regional changes in extremes are broadly consistent with the zonal means. In the Southern Hemisphere, the responses of both LWE (Figure 9) and HWE (Figure 10) generally track the poleward shift of the westerlies. Farther poleward, most models (except for TaiESM1; Supplementary Figures 1 and 5) show the HWE also intensifies over the Antarctica despite substantial zonal variations. In the Northern Hemisphere, the LWE responses over land and oceans are weak, and the signals related to a poleward-shifted jet are not evident in the responses of LWE and HWE (Supplementary Figures 1 and 2). Nonetheless, HWE intensifies up to ~1.0 m s$^{-1}$ over some land regions, the open ocean, and the Arctic. While the HWE intensification in the high-latitude regions shows notable regional and inter-model variations, its presence across models and surface boundary conditions (Aqua and AMIP poles) suggests the intensification is driven by warming-induced thermodynamical changes.

Notable spread in wind responses emerge in the tropical ocean owing to model structural uncertainty. Despite weak zonal mean responses (Figure 7), the weakening of regional wind is widespread over tropical oceans. Comparing Figures 8-10 suggests the weakening is limited for weak winds (Figure 9) and is mostly associated with changes in strong winds like HWE (Figure 10). Despite the widespread weakening tendency, its precise pattern and intensity are highly sensitive to individual model formulations. The most profound disagreements occur in the response of tropical HWE, where models differ not only in magnitude but also in the sign of the change in regions including the tropical Northwestern Pacific and North Indian Ocean (Figure 10). We acknowledge that CFMIP simulations may not accurately represent HWE due to the relatively low model resolution. Whether such differences exist in high-resolution climate simulations warrants future research.

Over land regions, the responses of mean winds and HWE show notable spread among the models. For example, CESM2 suggests a strong HWE response in South America, while CNRM and HadGEM show more pronounced changes in Central North America. Since the SST and other climate forcings are identical for each model, these large model differences likely arise from the differences in representing land-atmosphere interactions. This is confirmed by additional examination of terrestrial wind in the 4×$CO_2$ experiments with fixed SSTs, which generate responses like those in P4K and Future4K experiments (not shown). The comparison between CESM2 and TaiESM1 reinforces this finding (Supplementary Figure 5). Despite sharing a common lineage and simulating broadly similar responses over open ocean, CESM2 and TaiESM1 simulate different responses over Eurasia and other terrestrial regions, implicating surface process parameterizations as a key source of uncertainty. We speculate that the differences arise from some CESM2 updates, including updated subgrid orographic drag parameterization and Community Land Model (Danabasoglu et al. 2020). The sources of model spreads over land regions warrant future investigation.

The pronounced regional discrepancies underscore the pivotal role of land-atmosphere interactions in modulating extreme wind responses over terrestrial areas. Mechanistically, this uncertainty stems from the diversity of land model formulations, particularly regarding how bulk formulae parameterize surface fluxes and vertical gradients that directly determine 10-meter winds. Furthermore, differences in grid resolution alter the representation of topography and associated surface drag. Collectively, wind projections over land remain among the most uncertain, posing a significant challenge for climate impact assessments.

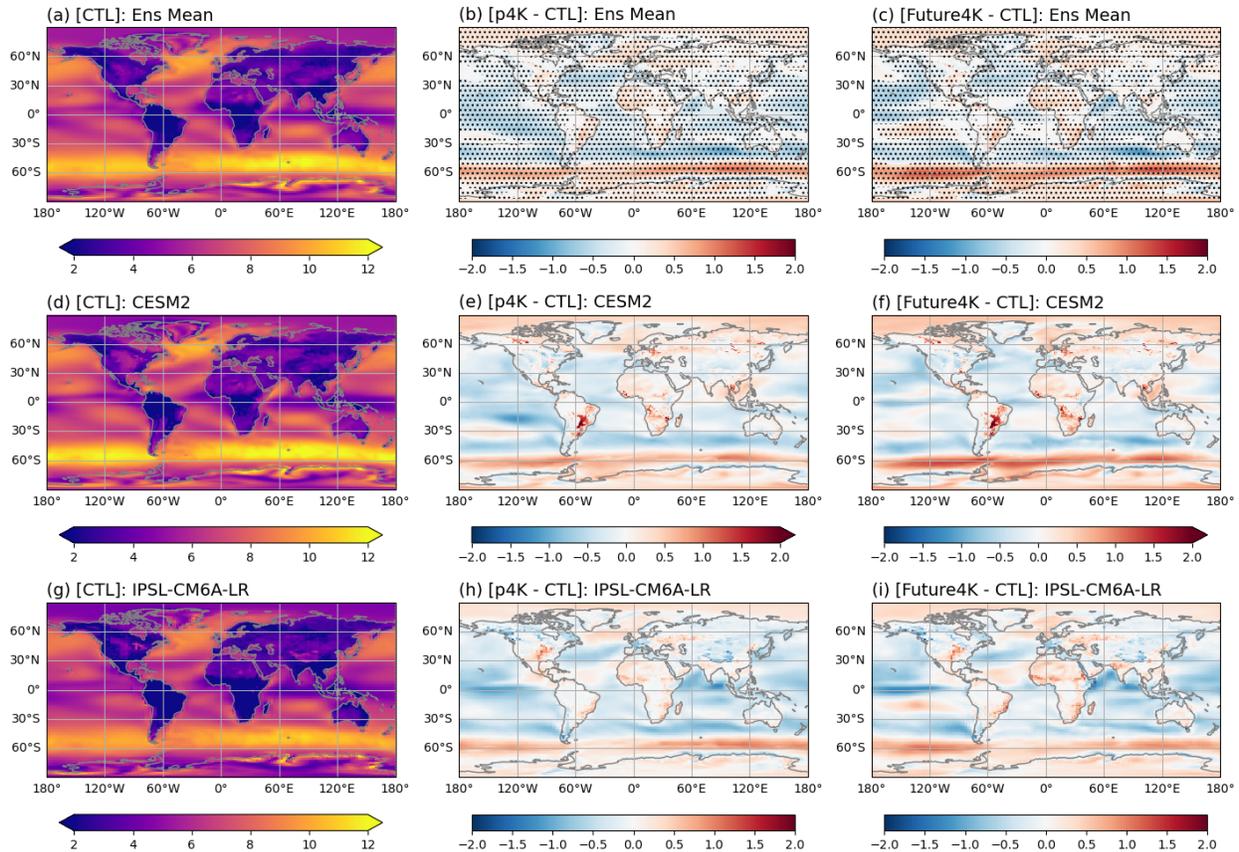

Figure 8 Mean surface wind speed and its responses to climate forcings in AMIP experiments. (a-f) Mean 10-meter wind speed in the AMIP pi-control experiments of multi-model mean (CESM2, CNRM-CM6-1, HadGEM3-GC31-LL, IPSL-CM6A-LR, MPI-ESM2-0), CESM2, and IPSL-CM6A-LR, respectively. (g-l) Same as (a-f), but for the differences between AMIP P4K and pi-control experiments. (m-r) Same as (a-f), but for the differences between AMIP Future4K and pi-control experiments. The stippling in (b-c) indicates regions where 80% models agree on the sign of changes. TaiESM1 does not provide Future4K daily outputs and is excluded.

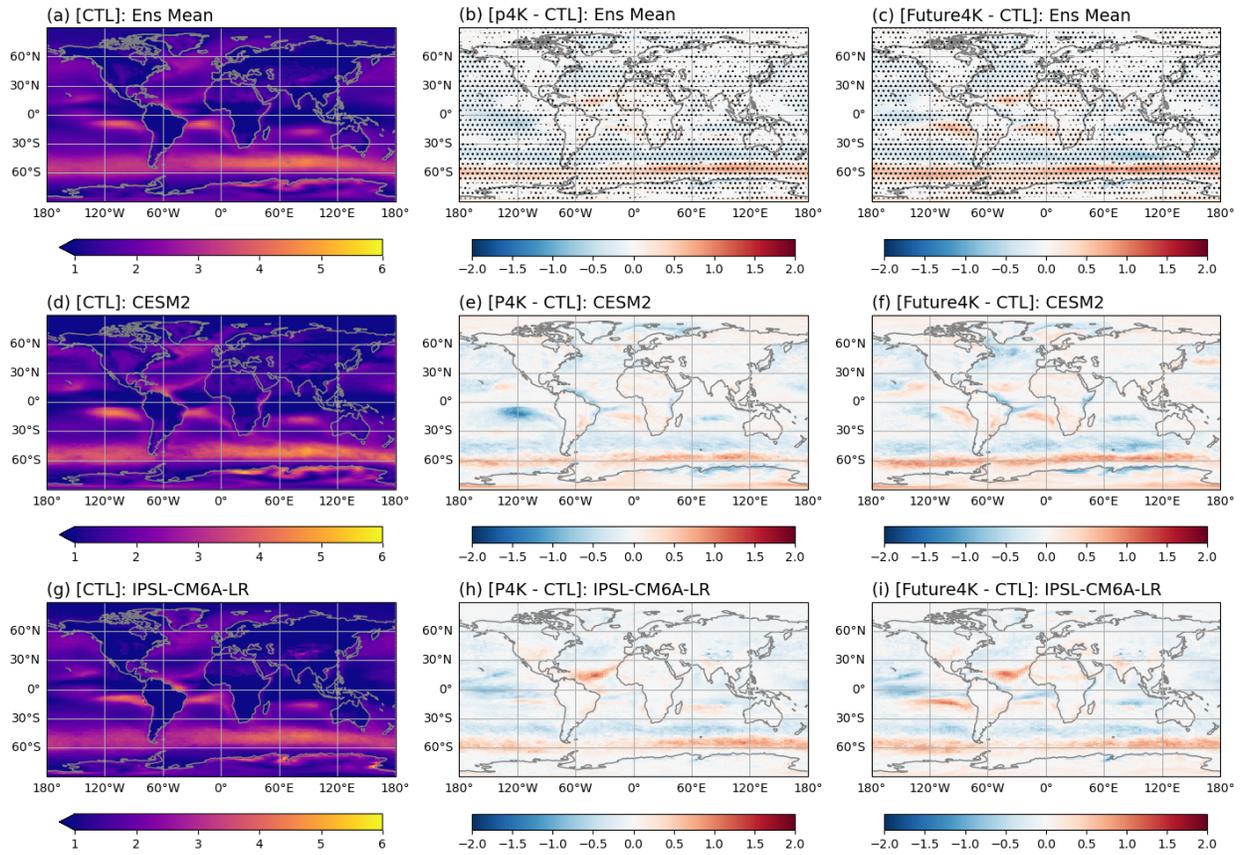

Figure 9 LWE climatology and its responses to climate forcings in AMIP experiments. The other settings are the same as Figure 8. TaiESM1 does not provide Future4K daily outputs and is excluded.

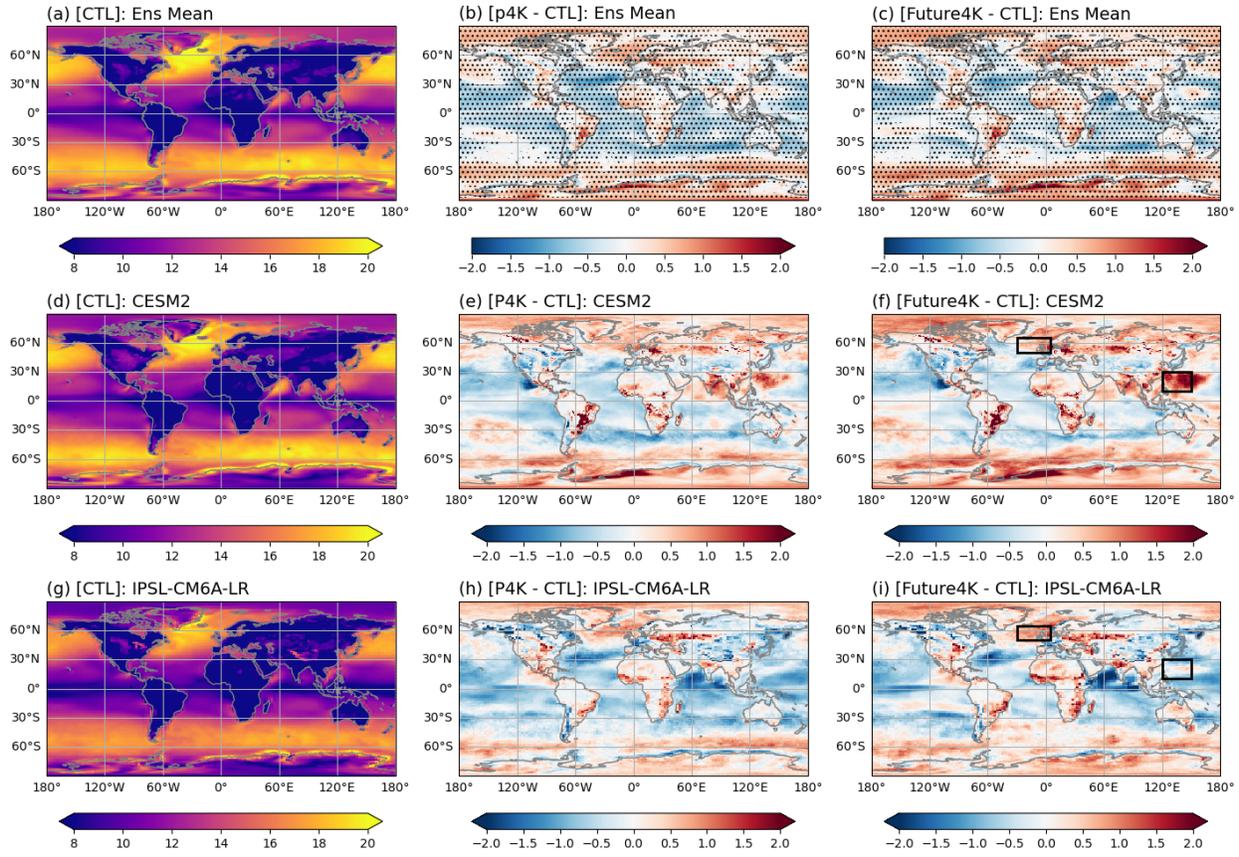

*Figure 10 HWE climatology and its responses to climate forcings in AMIP experiments. The black boxes in (f) and (i) denote the regions analyzed in Figures 11 and 12. The other settings are the same as Figure 8. TaiESM1 does not provide Future4K daily outputs and is excluded.*

Beyond land-atmosphere interactions, inter-model spread also stems from fundamental differences in how models represent the extreme-producing weather systems themselves. The disagreement about HWE responses in the Northwestern Pacific provides a stark example (Figure 11). Specifically, CESM2 suggests HWE strengthens in this region, whereas most other models (e.g., IPSL-CM6A-LR) suggest HWE weakens. Composites of the HWE events in CESM2 suggest that the simulated HWE strengthening can be traced to the strengthening of summertime tropical low-pressure systems near the edge of the subtropical high (Figures 11a, b). The synoptic-scale driver of HWE in the CESM2 AMIP experiments thus differs from the anticyclones in the CESM2 Aqua experiments (Figure 4g). In comparison, the HWE events in IPSL-CM6A-LR appear associated with wintertime anticyclones (Figures 11c, d). This synoptic-scale pattern resemble the Aqua simulations (Figures 4g and 6g) and is consistent with the weak tropical wave activity simulated by IPSL-CM6A-LR (Medeiros et al. 2021). Conversely, the projections are far more robust where models agree on the extreme-generating weather systems. In the North Atlantic, for instance, models consistently attribute HWE to wintertime extratropical cyclones (Figure 12). Consequently, despite differences in model formulation, the simulations show a robust intensification of HWE in this region, driven by the consistent dynamical response of these cyclones to warming. This contrast between the Pacific and Atlantic highlights that reducing regional uncertainty requires first ensuring that models simulate the correct season and nature of the extreme-generating weather systems.

Amidst these significant regional uncertainties, an intriguing and important point of robustness emerges: the large-scale patterns of extreme wind change are remarkably insensitive to the spatial pattern of SST warming. The uniform (P4K) and patterned (Future4K) warming experiments yield broadly similar responses in AMIP simulations. This suggests that for many large-scale aspects of extreme wind changes, the total magnitude of warming (which is comparable between P4K and Future4K experiments) is more influential than the specific warming pattern driven by anthropogenic forcings. This finding is particularly relevant given the uncertainties in future regional SST changes, especially in the tropical Pacific (Coats and Karnauskas 2017; Seager et al. 2022; Wills et al. 2022), and their potential impacts on changes in global extreme events (Sobel et al. 2023). While more complex or uncertain changes in tropical SSTs might elevate the importance of the warming pattern, our current analysis suggests that the overall warming effect is a primary driver, at least in the examined AMIP experiments with 4-K warming.

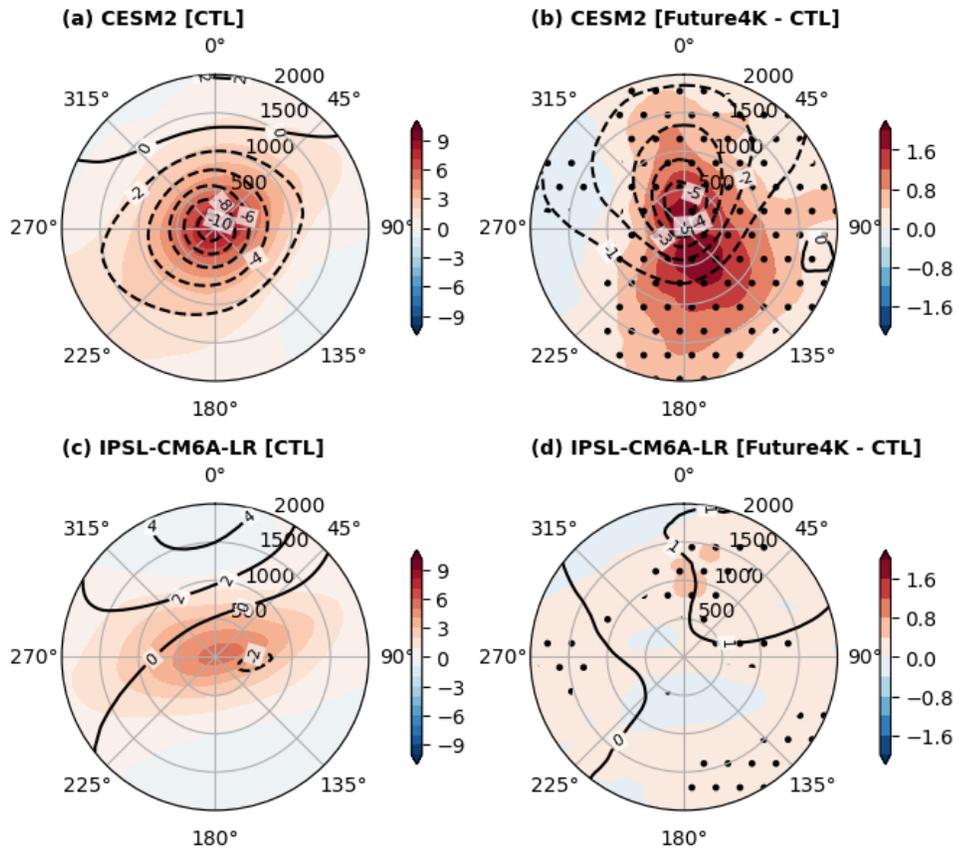

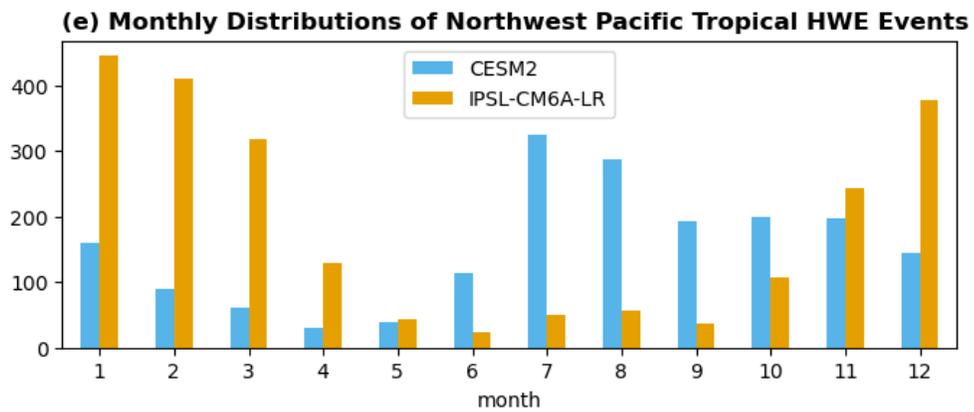

*Figure 11 Comparison of HWE events and their changes in the Northwest Pacific (10ºN-30ºN, 120ºE-150ºE). (a) Composites of HWE events in the CESM2 pi-control experiment. Color shading shows 10-meter wind anomalies (m s$^{-1}$), and contours show sea-level pressure anomalies (hPa). The stippling indicates 95% confidence level changes in sea level pressure. (b) Same as (a), but for the differences between CESM2 Future4K and pi-control experiments. (c-d) Same as (a-b), but for IPSL-CM6A-LR experiments. (e) Monthly counts of HWE events in the 30-year climatology of the AMIP simulation.*

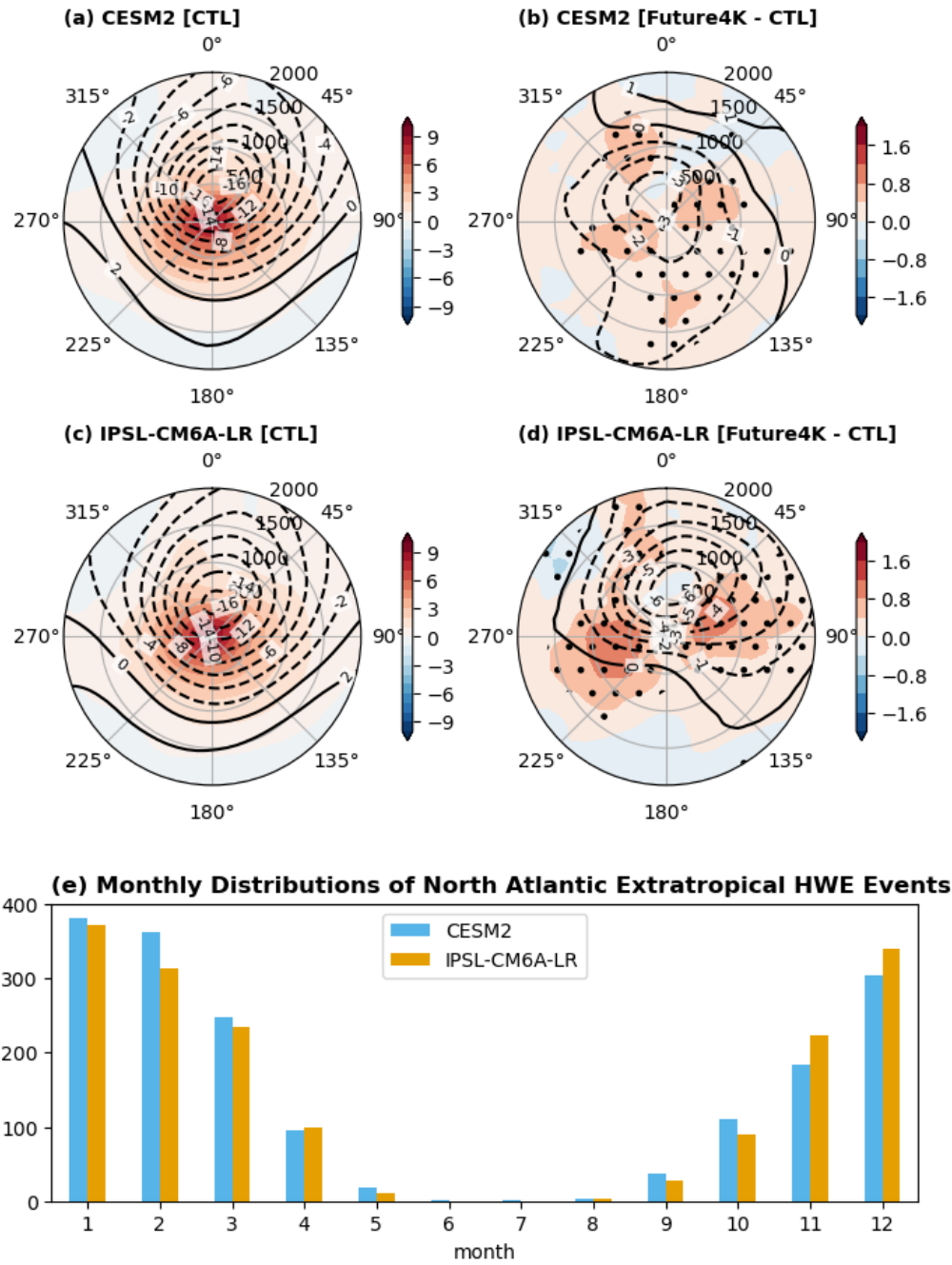

Figure 12 Comparison of HWE events and their changes in the high-latitude North Atlantic (50°N-65°N, 30°W-5°E). The other plot settings are the same as Figure 11.

# 5. Summary and Discussion

By systematically investigating near-surface extreme winds across a hierarchy of atmosphere-only models, this study answers the foundational questions posed earlier.

- **Impact of CO2 Forcing vs. SST Warming:** Direct $CO_2$ forcing produces weak responses, while SST warming strongly intensifies high wind extremes (HWE) at high latitudes. Under strong 4-K warming scenarios, the global magnitude of warming—rather than the specific regional SST pattern—dominates the large-scale changes in near-surface wind and its extremes.

- **Aqua vs. AMIP Representations:** Both setups show robust intensification of extratropical HWE, consistent with previous studies (Priestley and Catto 2022; Gentile et al. 2023). Realistic AMIP boundary conditions mute the Northern Hemisphere zonal-mean response and reduce inter-model spreads for tropical low wind extremes (LWE) compared to Aqua runs. Crucially, AMIP experiments expose land-atmosphere interactions as a major source of uncertainty of wind changes over land.

- **Links to Synoptic-Scale Weather Systems:** Regional uncertainty in wind extremes is directly tied to how models resolve extreme-producing weather systems. In the AMIP simulations, projections are robust where models agree on the driving dynamics (e.g., HWE in the extratropical North Atlantic), but diverge sharply when models disagree on the fundamental seasonality and type of the underlying synoptic systems (e.g., HWE in the tropical Northwestern Pacific).

While the overall warming magnitude may set the large-scale circulation response in future scenarios, recent observed trends and regional projections are sensitive to the precise SST pattern.

For example, the failure of coupled models to capture the observed recent SST trend pattern is a primary driver of model-reanalysis discrepancies for wintertime low-level extreme wind trends (Chou et al. 2025). This highlights that coupled model biases in projecting SST patterns are a dominant source of real-world uncertainty, a factor not fully captured by our uncoupled AMIP framework. Additionally, the Future4K and uniform P4K warming experiments did not fully account for the potential uncertainty of SST changes driven by natural variability or model errors. If tropical SSTs were to evolve in a highly uncertain or different manner, the warming pattern could become a more important source of uncertainty, especially on the regional scale.

The idealized Aqua experiments proved invaluable for isolating fundamental dynamics, revealing a stark contrast between the robust response of HWE at high latitudes and the highly uncertain response of LWE in the tropics. Our mechanistic analysis linked these changes to the differential response of weather systems. In the Aqua simulations, warming tends to intensify extratropical cyclones and strengthen HWE on the poleward flank of the midlatitude jet. The introduction of realistic boundary conditions in the AMIP simulations confirmed the robustness of the Southern Hemisphere extratropical response but also introduced new complexities. This included a muted Northern Hemisphere response and, counterintuitively, a reduction in inter-model disagreement in the tropics, suggesting that realistic boundary conditions provide a strong constraint on the flow.

Over land regions, the AMIP simulations highlighted the critical role of land-atmosphere interactions, which amplify regional responses and act as a key source of model structural uncertainty. Furthermore, wind changes over the land surface are also subject to impacts of SST changes, including the magnitude and pattern of warming. Additionally, the uncertainty of wind projections associated with land processes is not fully accounted for by our analysis, as aerosol forcings were removed or controlled in CFMIP simulations (Webb et al. 2017). The aerosol forcing can be important for understanding regional wind changes. For example, the recent summertime wind trends over Europe has been robustly attributed to aerosol forcings (Dong et al. 2022; Kang et al. 2024; Chou et al. 2025).

Several limitations warrant discussion. Our use of atmosphere-only or atmosphere-land models, while necessary for isolating the atmospheric response, omits crucial climate interactions involving the ocean and cryosphere. Future work using fully coupled models is essential to determine how these interactions might modulate the robust atmospheric responses identified here. Whether the lack of realistic atmosphere-ocean coupling, together with the specific warming pattern in Future4K experiments, affects the sensitivity of near-surface wind extremes to warming pattern warrants further investigation. Furthermore, the coarse resolution of the models examined limits their ability to resolve sharp-gradient phenomena, meaning our analysis of HWE (unlike LWE) probably is inadequate in capturing changes in meso-scale features (e.g., tropical cyclones and sting jets) or orography impacts (e.g., gap winds). Although high-resolution climate simulations (50-100 km grid spacing) show notable spreads (Zhang 2025), leveraging higher-resolution (e.g., convection-permitting) simulations represents a critical next step to understand changes in the most damaging wind events (Roberts et al. 2025).

Ultimately, this work demonstrates that reducing uncertainty in projections of societally relevant wind extremes requires a two-pronged approach. First, as our synoptic analysis shows, models must improve their fundamental internal physics—the "what" and "when" of extreme-producing weather systems. Second, models must also correctly simulate the external forcing and boundary conditions, as biases in coupled components like SST trend patterns can fundamentally distort some aspects of the large-scale environment and, consequently, the behavior of weather systems. It is valuable to focus on validating and improving both the fundamental physical representation of regional weather systems and the large-scale environment within our Earth system models.

# Acknowledgement

G.Z. thanks the organizers and participants of 2024 CFMIP meeting and Prof. Talia Tamarin-Brodsky for stimulating discussions that inspired this study. The research is supported by the U.S. National Science Foundation award AGS-2327959 and RISE-2530555, as well as the faculty development fund of the University of Illinois Urbana-Champaign. K.R. is supported by National Science Foundation award AGS-2327958 and this material is based upon work supported, in part, by the Department of Defense (DoD) Environmental Security Technology Certification Program (ESTCP) under Contract No. W912HQ25C0016. IRS and BM are supported by the NSF National Center for Atmospheric Research which is a major facility sponsored by the NSF under cooperative agreement No. 1852977.

# References

Andrews, M. B., and Coauthors, 2020: Historical Simulations With HadGEM3-GC3.1 for CMIP6. *J Adv Model Earth Syst*, **12**, e2019MS001995, https://doi.org/10.1029/2019MS001995.

Boucher, O., and Coauthors, 2020: Presentation and Evaluation of the IPSL-CM6A-LR Climate Model. *J Adv Model Earth Syst*, **12**, e2019MS002010, https://doi.org/10.1029/2019MS002010.

Chadwick, R., and Coauthors, 2026: Processes Controlling the South American Monsoon Response to Climate Change. *Journal of Climate*, **39**, 261–280, https://doi.org/10.1175/JCLI-D-25-0012.1.

Chou, H.-H., T. A. Shaw, and G. Zhang, 2025: Human influence on recent trends in extratropical low-level wind speed. *npj Clim Atmos Sci*, **9**, 20, https://doi.org/10.1038/s41612-025-01292-6.

Coats, S., and K. B. Karnauskas, 2017: Are Simulated and Observed Twentieth Century Tropical Pacific Sea Surface Temperature Trends Significant Relative to Internal Variability? *Geophysical Research Letters*, **44**, 9928–9937, https://doi.org/10.1002/2017GL074622.

Coumou, D., J. Lehmann, and J. Beckmann, 2015: The weakening summer circulation in the Northern Hemisphere mid-latitudes. *Science*, **348**, 324–327, https://doi.org/10.1126/science.1261768.

Dai, P., and J. Nie, 2022: Robust Expansion of Extreme Midlatitude Storms Under Global Warming. *Geophysical Research Letters*, **49**, e2022GL099007, https://doi.org/10.1029/2022GL099007.

Danabasoglu, G., and Coauthors, 2020: The Community Earth System Model Version 2 (CESM2). *J Adv Model Earth Syst*, **12**, e2019MS001916, https://doi.org/10.1029/2019MS001916.

Deser, C., and Coauthors, 2020: Insights from Earth system model initial-condition large ensembles and future prospects. *Nat. Clim. Chang.*, **10**, 277–286, https://doi.org/10.1038/s41558-020-0731-2.

Di Girolamo, L., G. Zhao, G. Zhang, Z. Wang, J. Loveridge, and A. Mitra, 2025: Decadal changes in atmospheric circulation detected in cloud motion vectors. *Nature*, **643**, 983–987, https://doi.org/10.1038/s41586-025-09242-1.

Dong, B., R. T. Sutton, L. Shaffrey, and B. Harvey, 2022: Recent decadal weakening of the summer Eurasian westerly jet attributable to anthropogenic aerosol emissions. *Nat Commun*, **13**, 1148, https://doi.org/10.1038/s41467-022-28816-5.

Elminir, H. K., 2005: Dependence of urban air pollutants on meteorology. *Science of The Total Environment*, **350**, 225–237, https://doi.org/10.1016/j.scitotenv.2005.01.043.

Gastineau, G., and B. J. Soden, 2009: Model projected changes of extreme wind events in response to global warming. *Geophysical Research Letters*, **36**, 2009GL037500, https://doi.org/10.1029/2009GL037500.

Gentile, E. S., M. Zhao, and K. Hodges, 2023: Poleward intensification of midlatitude extreme winds under warmer climate. *npj Clim Atmos Sci*, **6**, 219, https://doi.org/10.1038/s41612-023-00540-x.

Hadas, O., and Y. Kaspi, 2025: A Lagrangian Perspective on the Growth of Midlatitude Storms. *AGU Advances*, **6**, e2024AV001555, https://doi.org/10.1029/2024AV001555.

Held, I. M., 2005: The Gap between Simulation and Understanding in Climate Modeling. *Bull. Amer. Meteor. Soc.*, **86**, 1609–1614, https://doi.org/10.1175/BAMS-86-11-1609.

——, and B. J. Soden, 2006: Robust Responses of the Hydrological Cycle to Global Warming. *Journal of Climate*, **19**, 5686–5699, https://doi.org/10.1175/JCLI3990.1.

Kang, J. M., T. A. Shaw, and L. Sun, 2024: Anthropogenic Aerosols Have Significantly Weakened the Regional Summertime Circulation in the Northern Hemisphere During the Satellite Era. *AGU Advances*, **5**, e2024AV001318, https://doi.org/10.1029/2024AV001318.

Kuma, P., F. A. -M. Bender, and A. R. Jönsson, 2023: Climate Model Code Genealogy and Its Relation to Climate Feedbacks and Sensitivity. *J Adv Model Earth Syst*, **15**, e2022MS003588, https://doi.org/10.1029/2022MS003588.

Lee, C.-Y., S. J. Camargo, A. H. Sobel, and M. K. Tippett, 2020: Statistical–Dynamical Downscaling Projections of Tropical Cyclone Activity in a Warming Climate: Two Diverging Genesis Scenarios. *Journal of Climate*, **33**, 4815–4834, https://doi.org/10.1175/JCLI-D-19-0452.1.

Lionello, P., R. D'Agostino, D. Ferreira, H. Nguyen, and M. S. Singh, 2024: The Hadley circulation in a changing climate. *Ann NY Acad Sci*, **1534**, 69–93, https://doi.org/10.1111/nyas.15114.

Lorenz, D. J., and E. T. DeWeaver, 2007: Tropopause height and zonal wind response to global warming in the IPCC scenario integrations. *J. Geophys. Res.*, **112**, 2006JD008087, https://doi.org/10.1029/2006JD008087.

Ma, J., and S.-P. Xie, 2013: Regional Patterns of Sea Surface Temperature Change: A Source of Uncertainty in Future Projections of Precipitation and Atmospheric Circulation*. *Journal of Climate*, **26**, 2482–2501, https://doi.org/10.1175/JCLI-D-12-00283.1.

Maher, P., and Coauthors, 2019: Model Hierarchies for Understanding Atmospheric Circulation. *Reviews of Geophysics*, **57**, 250–280, https://doi.org/10.1029/2018RG000607.

Medeiros, B., B. Stevens, and S. Bony, 2015: Using aquaplanets to understand the robust responses of comprehensive climate models to forcing. *Clim Dyn*, **44**, 1957–1977, https://doi.org/10.1007/s00382-014-2138-0.

——, A. C. Clement, J. J. Benedict, and B. Zhang, 2021: Investigating the impact of cloud-radiative feedbacks on tropical precipitation extremes. *npj Clim Atmos Sci*, **4**, 18, https://doi.org/10.1038/s41612-021-00174-x.

Narinesingh, V., H. Guo, S. T. Garner, and Y. Ming, 2024: Uniform SST Warming Explains Most of the NH Winter Circulation and Blocking Response in a Warmer Climate. *Journal of Climate*, **37**, 4595–4612, https://doi.org/10.1175/JCLI-D-23-0371.1.

Pfahl, S., P. A. O'Gorman, and M. S. Singh, 2015: Extratropical Cyclones in Idealized Simulations of Changed Climates. *Journal of Climate*, **28**, 9373–9392, https://doi.org/10.1175/JCLI-D-14-00816.1.

Priestley, M. D. K., and J. L. Catto, 2022: Future changes in the extratropical storm tracks and cyclone intensity, wind speed, and structure. *Weather Clim. Dynam.*, **3**, 337–360, https://doi.org/10.5194/wcd-3-337-2022.

Roberts, M. J., and Coauthors, 2025: High-Resolution Model Intercomparison Project phase 2 (HighResMIP2) towards CMIP7. *Geosci. Model Dev.*, **18**, 1307–1332, https://doi.org/10.5194/gmd-18-1307-2025.

Schemm, S., and M. Röthlisberger, 2024: Aquaplanet simulations with winter and summer hemispheres: model setup and circulation response to warming. *Weather Clim. Dynam.*, **5**, 43–63, https://doi.org/10.5194/wcd-5-43-2024.

Seager, R., N. Henderson, and M. Cane, 2022: Persistent Discrepancies between Observed and Modeled Trends in the Tropical Pacific Ocean. *Journal of Climate*, **35**, 4571–4584, https://doi.org/10.1175/JCLI-D-21-0648.1.

Sharmila, S., and K. J. E. Walsh, 2018: Recent poleward shift of tropical cyclone formation linked to Hadley cell expansion. *Nature Clim Change*, **8**, 730–736, https://doi.org/10.1038/s41558-018-0227-5.

Shaw, T. A., 2019: Mechanisms of Future Predicted Changes in the Zonal Mean Mid-Latitude Circulation. *Curr Clim Change Rep*, **5**, 345–357, https://doi.org/10.1007/s40641-019-00145-8.

Shaw, T. A., and A. Voigt, 2015: Tug of war on summertime circulation between radiative forcing and sea surface warming. *Nature Geosci*, **8**, 560–566, https://doi.org/10.1038/ngeo2449.

Shaw, T. A., and O. Miyawaki, 2024: Fast upper-level jet stream winds get faster under climate change. *Nat. Clim. Chang.*, **14**, 61–67, https://doi.org/10.1038/s41558-023-01884-1.

Shaw, T. A., and Coauthors, 2024: Emerging Climate Change Signals in Atmospheric Circulation. *AGU Advances*, **5**, e2024AV001297, https://doi.org/10.1029/2024AV001297.

Sherwood, S. C., and Coauthors, 2020: An Assessment of Earth's Climate Sensitivity Using Multiple Lines of Evidence. *Reviews of Geophysics*, **58**, e2019RG000678, https://doi.org/10.1029/2019RG000678.

Shrestha, S., and B. J. Soden, 2023: Anthropogenic Weakening of the Atmospheric Circulation During the Satellite Era. *Geophysical Research Letters*, **50**, e2023GL104784, https://doi.org/10.1029/2023GL104784.

Simpson, I. R., T. A. Shaw, and R. Seager, 2014: A Diagnosis of the Seasonally and Longitudinally Varying Midlatitude Circulation Response to Global Warming. *Journal of the Atmospheric Sciences*, **71**, 2489–2515, https://doi.org/10.1175/JAS-D-13-0325.1.

——, and Coauthors, 2025: Confronting Earth System Model trends with observations. *Sci. Adv.*, **11**, eadt8035, https://doi.org/10.1126/sciadv.adt8035.

Sobel, A. H., and Coauthors, 2023: Near-term tropical cyclone risk and coupled Earth system model biases. *Proc. Natl. Acad. Sci. U.S.A.*, **120**, e2209631120, https://doi.org/10.1073/pnas.2209631120.

Stevens, B., and S. Bony, 2013: What Are Climate Models Missing? *Science*, **340**, 1053–1054, https://doi.org/10.1126/science.1237554.

Tamarin, T., and Y. Kaspi, 2017: The poleward shift of storm tracks under global warming: A Lagrangian perspective. *Geophysical Research Letters*, **44**, https://doi.org/10.1002/2017GL073633.

Tierney, G., D. J. Posselt, and J. F. Booth, 2018: An examination of extratropical cyclone response to changes in baroclinicity and temperature in an idealized environment. *Clim Dyn*, **51**, 3829–3846, https://doi.org/10.1007/s00382-018-4115-5.

Ullrich, P. A., C. M. Zarzycki, E. E. McClenny, M. C. Pinheiro, A. M. Stansfield, and K. A. Reed, 2021: TempestExtremes v2.1: a community framework for feature detection, tracking, and analysis in large datasets. *Geosci. Model Dev.*, **14**, 5023–5048, https://doi.org/10.5194/gmd-14-5023-2021.

Voldoire, A., and Coauthors, 2019: Evaluation of CMIP6 DECK Experiments With CNRM-CM6-1. *J Adv Model Earth Syst*, **11**, 2177–2213, https://doi.org/10.1029/2019MS001683.

Webb, M. J., and Coauthors, 2017: The Cloud Feedback Model Intercomparison Project (CFMIP) contribution to CMIP6. *Geosci. Model Dev.*, **10**, 359–384, https://doi.org/10.5194/gmd-10-359-2017.

Wills, R. C. J., R. H. White, and X. J. Levine, 2019: Northern Hemisphere Stationary Waves in a Changing Climate. *Curr Clim Change Rep*, **5**, 372–389, https://doi.org/10.1007/s40641-019-00147-6.

——, Y. Dong, C. Proistosecu, K. C. Armour, and D. S. Battisti, 2022: Systematic Climate Model Biases in the Large-Scale Patterns of Recent Sea-Surface Temperature and Sea-Level Pressure Change. *Geophysical Research Letters*, **49**, e2022GL100011, https://doi.org/10.1029/2022GL100011.

Woollings, T., M. Drouard, C. H. O'Reilly, D. M. H. Sexton, and C. McSweeney, 2023: Trends in the atmospheric jet streams are emerging in observations and could be linked to tropical warming. *Commun Earth Environ*, **4**, 125, https://doi.org/10.1038/s43247-023-00792-8.

Yin, J. H., 2005: A consistent poleward shift of the storm tracks in simulations of 21st century climate. *Geophysical Research Letters*, **32**, 2005GL023684, https://doi.org/10.1029/2005GL023684.

Yukimoto, S., and Coauthors, 2019: The Meteorological Research Institute Earth System Model Version 2.0, MRI-ESM2.0: Description and Basic Evaluation of the Physical Component. *Journal of the Meteorological Society of Japan*, **97**, 931–965, https://doi.org/10.2151/jmsj.2019-051.

Zeng, Z., and Coauthors, 2019: A reversal in global terrestrial stilling and its implications for wind energy production. *Nat. Clim. Chang.*, **9**, 979–985, https://doi.org/10.1038/s41558-019-0622-6.

Zhang, G., 2023: Warming-induced contraction of tropical convection delays and reduces tropical cyclone formation. *Nat Commun*, **14**, 6274, https://doi.org/10.1038/s41467-023-41911-5.

——, 2025: Amplified summer wind stilling and land warming compound energy risks in Northern Midlatitudes. *Environ. Res. Lett.*, **20**, 034015, https://doi.org/10.1088/1748-9326/adb1f8.

Zhao, M., and T. Knutson, 2024: Crucial role of sea surface temperature warming patterns in near-term high-impact weather and climate projection. *npj Clim Atmos Sci*, **7**, 130, https://doi.org/10.1038/s41612-024-00681-7.